\documentclass[oldversion]{aa} 

\usepackage{graphicx}
\usepackage{txfonts}
\begin{document}
 
\title{
Strong Lensing by a Node of the Cosmic Web}
   \titlerunning{Strong Lensing by a Node of the Cosmic Web}
   \authorrunning{Limousin et~al.}
   \subtitle{The Core of MACS\,J0717.5+3745 at $z=0.55$}
   \author{M. Limousin\inst{1,2},
H. Ebeling\inst{3},
J. Richard\inst{4},
A.~M. Swinbank\inst{5},
G.~P. Smith\inst{6,7},
S. Rodionov\inst{1},
C.-J. Ma\inst{8,9},
I. Smail\inst{5},\\
A.~C. Edge\inst{5},
M. Jauzac\inst{1},
E. Jullo\inst{10,1}
\& J.-P. Kneib\inst{1}
      \thanks{Based on observations obtained with the \emph{Hubble Space Telescope}
	and the Keck Telescope.
       }
       }
   \offprints{marceau.limousin@oamp.fr}

   \institute{
	Laboratoire d'Astrophysique de Marseille, UMR\,6610, CNRS-Universit\'e de Provence,
	38 rue Fr\'ed\'eric Joliot-Curie, 13\,388 Marseille Cedex 13, France
	\and
         Dark Cosmology Centre, Niels Bohr Institute, University of Copenhagen,
        Juliane Maries Vej 30, 2100 Copenhagen, Denmark
	\and
	Institute for Astronomy, University of Hawaii, 2680 Woodlawn Dr, Honolulu, HI 96822, USA
	\and
	CRAL, Observatoire de Lyon, Université Lyon 1, 9 Avenue Ch. André, 69561 Saint Genis Laval Cedex, France
	\and
	Institute for Computational Cosmology, Department of Physics, Durham University, South Road, Durham DH1 3LE, UK
	\and
	School of Physics and Astronomy, University of Birmingham, Edgbaston, Birmingham B15 2TT
	\and
	Department of Astronomy, California Institute of Technology, 105-24, Pasadena, CA91125, USA
	\and
	Department of Physics \& Astronomy, University of Waterloo, 200 University Ave. W., Waterloo, Ontario, N2L 3G1, Canada
	\and
	Harvard-Smithsonian Center for Astrophysics, 60 Garden St., Cambridge, Massachusetts, 02138-1516, USA
	\and
	Jet Propulsion Laboratory, Caltech, MS 169-327, 4\,800 Oak Grove Dr, Pasadena, CA 91109, USA
              }

   
  \abstract
   {
	We present results of a strong-lensing analysis of MACS\,J0717.5+3745 
(hereafter MACS\,J0717), an extremely X-ray luminous galaxy cluster at
$z=0.55$. Observations at different wavelengths
reveal a complex and dynamically very active cluster, whose core is connected to a
large scale filament extended over several Mpc.
Using multi-passband imaging data obtained with the {\it Hubble Space Telescope}'s Advanced Camera for Surveys (ACS), we identify 15 multiply imaged systems across
the full field of view of ACS, five of which we confirmed 
spectroscopically in groundbased follow-up observations with the Keck telescope.
We use these multiply imaged systems to constrain a parametric model of the mass distribution in the cluster
core, employing a new parallelized version of the \textsc{Lenstool} software. The main result is that the most
probable description of the mass distribution comprises four cluster-scale
dark matter haloes. The total mass distribution follows the light distribution but strongly deviates from 
the distribution of the intra-cluster gas as traced by the X-ray surface brightness.
This confirms the complex morphology proposed by previous studies.
We interpret this segregation of collisional and collisionless matter as strong evidence of multiple mergers and ongoing dynamical activity.
MACS\,J0717 thus constitutes one of the most disturbed clusters presently known and, featuring
a projected mass within the ACS field of view (\,R\,=\,150$\arcsec$=\,960\,kpc\,) of 
2.11$\pm$0.23 $\times$ 10$^{15}$ M$_{\sun}$, the system is also one of the most massive known.
   }

   \keywords{Gravitational lensing: strong  --
               Galaxies: clusters: individual: MACS\,J0717.5+3745
	     }

   \maketitle

\end{document}